\documentstyle[twoside,psfig]{adacta}

\begin{document}

\unitlength1mm
\prvastrana=1
\poslednastrana=10

\def\autor{S. Scheel, L. Kn\"oll, D.-G. Welsch}
\def\nazov{Spontaneous decay in the presence
of absorbing dielectric bodies}

\headings{1}{10}

\title{SPONTANEOUS DECAY IN THE PRESENCE OF 
ABSORBING DIELECTRIC BODIES}

\author{S. Scheel, L. Kn\"oll, 
D.-G.Welsch\footnote{\email{welsch@tpi.uni-jena.de}}}
{Theoretisch-Physikalisches Institut, Friedrich-Schiller-Universit\"{a}t 
Jena, Max-Wien-Platz 1, D-07743 Jena, Germany}

\datumy{23 April 1999}{}

\abstract{We present a formalism for studying the influence of
dispersive and absorbing dielectric bodies on a radiating atom
in the framework of quantization of the phenomenological 
Maxwell equations for given complex permittivities of the bodies.   
In Markov approximation, the rate of spontaneous decay and the 
line shift associated with it can then be related to the
complex permittivities and geometries of the bodies via the 
dyadic Green function of the classical boundary value problem 
of electrodynamics -- a result which is in agreement 
with second-order calculations for microscopic model systems.
The theory is applied to an atom near a planar interface as well as 
to an atom in a spherical cavity. The latter, also known as the 
real-cavity model for spontaneous decay of an excited atom embedded 
in a dielectric, is compared with the virtual-cavity model. 
Connections with other approaches are mentioned and 
the results are compared.}

%%%%%%%%%%%%%%%%%%%%%%%%%%%%%%%%%%%%%%%%%%%%%%%%%%%%%%%%%%%%%%%%%%%%%%

\section{Introduction}
\label{intro}

It is well known that the quantum statistics of photons and atoms 
can essentially be influenced by the presence of macroscopic
bodies. In principle, such bodies can be included in the
quantum mechanical description of the system as a part of matter 
to which the radiation field is coupled and treated microscopically,
using, e.g., a canonical Hamiltonian formalism and
perturbation theory (mostly up to second order). Apart from the 
fact that in practice the involved calculations 
can only be performed for relatively simple systems, each
system requires its own calculation in general, because of the
specific microscopic properties that must be taken into account. 

In classical optics the effect of macroscopic bodies is
commonly described phenomenologically in terms of a spatially
varying permittivity $\epsilon({\bf r},\omega)$
[or refractive index $n({\bf r},\omega)$ $\!=$
$\sqrt{\epsilon({\bf r},\omega)}$] which is a complex 
function of frequency in general. Recently a scheme for 
transferring this concept to quantum theory has been
developed [1,2,3] which is consistent with 
standard QED as well as with the dissipation-fluctuation theorem. 
The advantage of such an approach is that it is valid for arbitrary 
linear dielectrics. Knowledge is needed only about the 
dispersion-absorption profile of the medium expressed in
terms of the (measurable) permittivity, from which the
dyadic Green function of the classical problem can be
calculated. The electric and magnetic fields can then
be expressed in terms of the classical Green function and
a continuum of fundamental bosonic fields.

In the present paper we use that quantization scheme for studying the 
problem of a radiating atom in the presence of dispersive and
absorbing dielectric matter. In particular, we consider
the effect on the spontaneous decay of an excited atom 
and calculate the decay rate and the line shift. With
regard to potential applications in scanning near-field optical
microscopy, we present results for an atom near a planar
interface. Further, we derive a correct generalization for
absorbing media of the real-cavity model and the virtual-cavity model
of spontaneous decay in dielectrics. 

The paper is organized as follows. In Section \ref{basic}
the quantization scheme for the electromagnetic field in linear
Kramers-Kronig dielectrics is briefly reviewed, and
general formulas for the decay rate and the line shift of 
an atom in the vicinity of a dielectric body are derived. 
In Section \ref{planar} a planar interface is 
considered, and Section \ref{real} is devoted to a spherical 
cavity, which may serve as a model for 
spontaneous decay in media. The corresponding
virtual-cavity model is considered in Section \ref{virtual},
and a comparison with the real-cavity model is given.
Finally, a summary is given in Section \ref{summary}.

%%%%%%%%%%%%%%%%%%%%%%%%%%%%%%%%%%%%%%%%%%%%%%%%%%%%%%%%%%%%%%%%%%%%%%

\section{Basic equations}
\label{basic}

Our analysis is based on the phenomenological quantization of the
electromagnetic field in linear Kramers-Kronig dielectrics as developed in
[1,2,3]. Without external sources, Maxwell's equations
in (temporal) Fourier space read
\begin{equation} \label{2.1}
{\bf \nabla} \cdot \underline{\hat{\bf B}}({\bf r},\omega) = 0,
\end{equation}
\begin{equation} \label{2.2}
{\bf \nabla} \cdot \left[ \epsilon_0 \epsilon({\bf r},\omega)
\underline{\hat{\bf E}}({\bf r},\omega) \right] = \underline{\hat{\rho}}({\bf
r},\omega) ,
\end{equation}
\begin{equation} \label{2.3}
{\bf \nabla} \times \underline{\hat{\bf E}}({\bf r},\omega) = i\omega
{\bf \nabla} \times \underline{\hat{\bf B}}({\bf r},\omega) ,
\end{equation}
\begin{equation} \label{2.4}
{\bf \nabla} \times \underline{\hat{\bf B}}({\bf r},\omega) =
-i\frac{\omega}{c^2} \epsilon({\bf r},\omega) \underline{\hat{\bf
E}}({\bf r},\omega) +\mu_0 \underline{\hat{\bf j}}({\bf r},\omega) ,
\end{equation}
with $\epsilon({\bf r},\omega)$ $\!=$ $\!\epsilon_R({\bf r},\omega)$
$\!+$ $\!i\epsilon_I({\bf r},\omega)$
being the spatially varying, complex  
permittivity of the medium which satisfies the Kramers-Kronig relations.
In Eqs.~(\ref{2.2}) and (\ref{2.4}),
$\underline{\hat{\rho}}({\bf r},\omega)$ and 
$\underline{\hat{\bf j}}({\bf r},\omega)$, respectively, are the 
operator noise charge and current density operators that are 
associated with absorption according to the dissipation-fluctuation 
theorem. They are related to the noise polarization 
$\underline{\hat{\bf P}}{}{}^N({\bf r},\omega)$ as
\begin{eqnarray} \label{2.5}
\underline{\hat{\bf j}}({\bf r},\omega) &=& -i\omega
\underline{\hat{\bf P}}{}^N({\bf r},\omega), \\ \label{2.6}
\underline{\hat{\rho}}({\bf r},\omega) &=& - {\bf \nabla} \cdot
\underline{\hat{\bf P}}{}^N({\bf r},\omega),
\end{eqnarray}
and hence satisfy the equation of continuity. By introducing an
infinite set of bosonic field operators 
$\hat{\bf f}({\bf r},\omega)$, all electromagnetic field 
operators can be expressed in terms of them using the fundamental relation
\begin{equation} \label{2.7}
\underline{\hat{\bf j}}({\bf r},\omega) = \omega \sqrt{\frac{\hbar
\epsilon_0}{\pi} \epsilon_I({\bf r},\omega)} \,
\hat{\bf f}({\bf r},\omega) .
\end{equation}
In particular, Maxwell's equations imply that the electric field
operator may be written in terms of the classical dyadic Green
function as
\begin{equation} \label{2.8}
\underline{\hat{E}}_k({\bf r},\omega) = i\mu_0 \int d^3{\bf s} \,
G_{kk'}({\bf r},{\bf r}',\omega) \,\underline{\hat{\!j}}_{k'}({\bf r}',\omega),
\end{equation}
where $G_{kk'}({\bf r},{\bf s},\omega)$ satisfies the partial
differential equation
\begin{equation} \label{2.9}
\left[ \partial_i^r \partial_k^r -\delta_{ik} \left( \triangle^r
+\frac{\omega^2}{c^2} \,\epsilon({\bf r},\omega) \right) \right]
G_{kl}({\bf r},{\bf r}',\omega) = \delta_{il} \delta({\bf r}-{\bf r}') .
\end{equation}
Note that the relation
\begin{equation} \label{2.9a}
\int d^3{\bf s} \,\frac{\omega^2}{c^2} \, \epsilon_I({\bf s},\omega)
G_{kl}({\bf r},{\bf s},\omega) G_{k'l}^\ast({\bf r'},{\bf s},\omega) = {\rm
Im}\,G_{kk'}({\bf r},{\bf r'},\omega)
\end{equation}
is valid, which can be checked by adding the complex conjugate 
of Eq.~(\ref{2.9}) to itself and integrating by parts. 

Integrating Eq.~(\ref{2.8}) with respect to $\omega$ leads to the
operator of the electric field and, via Maxwell's equations, to the
magnetic field operator, with the equal-time commutation relations
\begin{equation} \label{2.10}
\epsilon_0 \left[ \hat{E}_k({\bf r}), \hat{B}_l({\bf r}') \right] = -i
\hbar \epsilon_{klm} \partial_m^r \delta({\bf r}-{\bf r}') ,
\end{equation}
\begin{equation}
\left[ \hat{E}_k({\bf r}), \hat{E}_l({\bf r}') \right] =
\left[ \hat{B}_k({\bf r}), \hat{B}_l({\bf r}') \right] = 0,
\end{equation}
which has been shown [3] to be valid for arbitrary linear
dielectrics that are consistent with the Kramers-Kronig relations.

If there is an additional (two-level) atom present, then the total
Hamiltonian of the system in dipole and rotating wave approximation
can be given by
\begin{equation} \label{2.11}
\hat{H} = \!\int\! d^3{\bf r}\! \int_0^\infty \!\!d\omega \,\hbar \omega
\hat{\bf f}^\dagger({\bf r},\omega)
\!\cdot\! \hat{\bf f}({\bf r},\omega)
+\!\sum\limits_{\alpha=1}^2 \hbar \omega_\alpha \hat{A}_{\alpha \alpha}
-\left[ i\omega_{21} \hat{A}_{21} \,\hat{\bf A}^{(+)}({\bf r}_A) \!\cdot\! 
{\bf d}_{21} +{\rm H.c.} \right] ,
\end{equation}
where we have chosen the Weyl gauge for the vector potential, and hence
the scalar potential is set equal to zero. It can then be shown 
that in Markov approximation 
the atomic (flip) operators $\hat{A}_{\alpha\alpha'}$ 
$\!=$ $\!|\alpha\rangle\langle\alpha'|$ satisfy the following 
Langevin-type differential equations [4]: 
\begin{eqnarray} \label{2.12}
\,\dot{\!\hat{A}}_{22} &=& -\Gamma \hat{A}_{22} -\left[ \hat{A}_{21}
\frac{\omega_{21}}{\hbar} \hat{\bf A}_{\rm free}^{(+)}({\bf r}_A,t)
\cdot {\bf d}_{21} +{\rm H.c.} \right] \\ \label{2.13}
\,\dot{\!\hat{A}}_{11} &=& -\,\dot{\!\hat{A}}_{22} \\ \label{2.14}
\,\dot{\!\hat{A}}_{21} &=& \left[ i(\omega_{21}- \delta\omega)
-{\textstyle\frac{1}{2}} \Gamma \right] \hat{A}_{21} 
+\frac{\omega_{21}}{\hbar} \hat{\bf A}_{\rm free}^{(-)}({\bf r}_A,t)
\cdot {\bf d}_{21} \left( \hat{A}_{22} -\hat{A}_{11} \right) ,
\end{eqnarray}
where the rate of spontaneous decay of the excited atomic state 
$|2\rangle$ is given by
\begin{equation} \label{2.15}
\Gamma = \frac{2\omega_A^2 \mu_k \mu_l}{\hbar \epsilon_0 c^2}\, 
{\rm Im}\,G_{kl}({\bf r}_A,{\bf r}_A,\omega_A),
\end{equation}
and the Lamb shift contribution reads
\begin{equation} \label{2.16}
\delta\omega = \frac{2\omega_A^2 \mu_k \mu_l}{\hbar \epsilon_0 c^2}
\left[ {\rm Re}\,G_{kl}({\bf r}_A,{\bf r}_A,\omega_A) -\frac{1}{\pi}
\int_0^\infty \!d\omega \, \frac{{\rm Im}\,G_{kl}({\bf r}_A,{\bf
r}_A,\omega)}{\omega+\omega_A} \right]
\end{equation}
[$\mu_k$ $\!\equiv$ $\!(d_{21})_k$, $\omega_A$ $\!\equiv$
$\!\omega_{21}$]. Note that the coincidence limit of the real 
part of the vacuum Green function is infinite and regularization
is required. Here and in the following the vacuum Lamb shift is
thought of as being included in the atomic transition frequency,
so that $\delta\omega$ only results from the medium.
With respect to Green functions of the type given in Eq.~(\ref{3.1}), with
$G^V_{kl}({\bf r}_A,{\bf r}_A,\omega_A)$ being the vacuum
Green function,
$G_{kl}({\bf r}_A,{\bf r}_A,\omega_A)$ in Eq.~(\ref{2.16})
can therefore be replaced with $R_{kl}({\bf r}_A,{\bf r}_A,\omega_A)$, and
Eq.~(\ref{2.15}) reduces to
\begin{equation} \label{2.16a}
\Gamma = \Gamma_0 +\frac{2\omega_A^2 \mu_k \mu_l}{\hbar \epsilon_0 c^2}\, 
{\rm Im}\,R_{kl}({\bf r}_A,{\bf r}_A,\omega_A),
\end{equation}
where $\Gamma_0$ $\!=$ $\!\omega_A^3\mu^2/(3\pi\hbar\epsilon_0c^3)$
is the spontaneous emission rate of the atom in free space.

Equations (\ref{2.15}) and (\ref{2.16}) hold for all
linear dielectrics and arbitrary geometries. Note, that
Eq.~(\ref{2.15}) can be rewritten, on using the
relation (\ref{2.9a}), as
\begin{equation} \label{2.17}
{\rm Im}\,G_{kk'}({\bf r},{\bf r'},\omega) \delta(\omega-\omega') =
\frac{\pi \epsilon_0 c^2}{\hbar \omega^2} \langle 0 | \left[
\underline{\hat{E}}_k({\bf r},\omega) ,
\underline{\hat{E}}_{k'}^\dagger({\bf r'},\omega') \right] | 0 \rangle ,
\end{equation}
which is nothing but a consequence of the consistent introduction of 
the noise polarization according to the dissipation-fluctuation
theorem. It should be pointed out that for an atom in vacuum the 
imaginary part of the Green function for arbitrary geometries 
of the surrounding dielectric is purely transverse and, most importantly, 
finite.

%%%%%%%%%%%%%%%%%%%%%%%%%%%%%%%%%%%%%%%%%%%%%%%%%%%%%%%%%%%%%%%%%%%%%%

\section{Planar interface}
\label{planar}

Let us first consider the spontaneous decay of an excited atom near 
an absorbing planar dielectric surface. This configuration with 
real refractive index  has been studied extensively in the literature 
in connection with Casimir and van der Waals forces (see [5] and 
references therein) as well as applications in scanning near-field optical 
microscopy (SNOM) (see [6] and references therein). To be
more specific, we consider two infinite half-spaces (volumes ${\cal V}_1$ 
and ${\cal V}_2$) with a common interface such that $\epsilon({\bf
r},\omega)$ $\!=$ $\!1$ if ${\bf r}$ $\!\in$ $\!{\cal V}_1$, and
$\epsilon({\bf r},\omega)$ $\!=$ $\!\epsilon(\omega)$ if
${\bf r}$ $\!\in$ $\!{\cal V}_2$
and assume that the atom is placed in ${\cal V}_1$ 
at a distance $z$ from the interface. The Green function 
in ${\cal V}_1$ for this configuration can be given by [2]
\begin{equation} \label{3.1}
G_{kk'}({\bf r},{\bf r'},\omega) = 
G_{kk'}^V({\bf r},{\bf r'},\omega) +R_{kk'}({\bf r},{\bf r'},\omega) 
\quad  ({\bf r},{\bf r'}\in {\cal V}_1), 
\end{equation}
where $G_{kk'}^V({\bf r},{\bf r'},\omega)$ is the vacuum
Green function, and $R_{kk'}({\bf r},{\bf r'},\omega)$ 
is the reflection Green function. 
As already mentioned, for calculating the decay rate
and the line shift we only need the reflection Green function with 
both spatial arguments at the position of the atom,
$R_{kk'}$ $\!\equiv$ $\!R_{kk'}({\bf r}_A,{\bf r}_A,\omega)$.
Using the formulas given in [2], we obtain
\begin{eqnarray} \label{3.2}
R_{xx} = R_{yy}
\hspace{-1.5ex}&=&\hspace{-1.5ex} -\frac{i}{8\pi q^2} \int_0^\infty \!dk \, k\beta \,
{\rm e}^{2i\beta z} r^p(k) +\frac{i}{8\pi} \int_0^\infty \!dk \,
\frac{k e^{2i\beta z}}{\beta} \, r^s(k), 
\\ \label{3.3}
R_{zz} \hspace{-1.5ex}&=&\hspace{-1.5ex} \frac{i}{4\pi q^2}
\int_0^\infty \!dk \, k^3 \,\frac{e^{2i\beta z}}{\beta} r^p(k)
\end{eqnarray}
($q$ $\!=$ $\!\omega/c$, $\beta$ $\!=$ $\!\sqrt{q^2-k^2}$). The
functions $r^p(k)$ and $r^s(k)$ are the usual Fresnel reflection
coefficients for $p$- and $s$-polarized waves. The integrals can be
evaluated asymptotically in the case when $qz$ $\!\ll$ $\!1$, i.e. when 
the distance of the atom from the surface is smaller than the transition
wavelength, which is in agreement with the Markov approximation. 
Expanding in powers of $z$ leads to [$\epsilon$ 
$\!\equiv$ $\!\epsilon(\omega)$,
$n$ $\!\equiv$ $\!\sqrt{\epsilon(\omega)}$]
\begin{eqnarray} \label{3.4}
R_{zz} \hspace{-1.5ex}&=&\hspace{-1.5ex} \frac{1}{16\pi q^2 z^3}
\frac{n^2-1}{n^2+1} +\frac{1}{8\pi z} \frac{(n-1)^2}{n(n+1)}
+\frac{iq}{12\pi} \frac{(n\!-\!1)(2n\!-\!1)}{n(n+1)} + {\cal O}(z), \\
\label{3.5} 
R_{xx} \hspace{-1.5ex}&=& \hspace{-1.5ex}R_{yy} = \frac{1}{2} R_{zz} 
-\frac{1}{16\pi z} \frac{n^2-1}{n^2+1} 
-\frac{iq}{3\pi} \frac{n-1}{n+1} +{\cal O}(z).
\end{eqnarray}
Inserting Eqs.~(\ref{3.4}) and (\ref{3.5}) into Eq.~(\ref{2.16a}) yields
\vspace{-0.5ex}
\begin{equation} \label{3.6}
\Gamma = \Gamma_0 
\left(1+\frac{\mu_z^2}{\mu^2}\right)\frac{3}{8(qz)^3}
\frac{\epsilon_I(\omega_A)}{|\epsilon(\omega_A)+1|^2} +{\cal
O}(z^{-1}) .
\end{equation}
The term proportional to $z^{-3}$ exactly agrees with the formula 
for the leading term derived in [7] in the framework of a 
microscopic approach, after lengthy and involved diagonalization 
of the model Hamiltonian used and computation of perturbation
series up to second order. It is worth noting that 
the phenomenologically introduced permittivity and
the Green function that is based on it already contain
the relevant information, so that the effort is reduced
drastically. Needless to say that the microscopic treatment 
given in [7] is far from a genuine {\em ab initio} calculation.

SNOM detects surface corrugation or impurities
via the changes in the line shift and the line width
of an atomic transition far from medium resonances [6].
The high sensitivity of the method results from the 
cubic dependence of the line shift on the inverse distance of the
probe atom from the surface. 
Disregarding absorption  %($\epsilon_I$ $\!\equiv$ $\!0$)
the line width is weakly distance-dependent 
[cf. Eqs.~(\ref{2.16a}) and (\ref{3.4})], and the
strong dependence on distance of the line shift
is essentially determined by the first term in Eq.~(\ref{2.16}), 
\vspace{-0.5ex}
\begin{equation} \label{3.6a}
\delta\omega = \Gamma_0 \left(1+\frac{\mu_z^2}{\mu^2}\right)
\frac{3}{4(qz)^3} \frac{\epsilon_R(\omega_A)-1}{\epsilon_R(\omega_A)+1} 
+{\cal O}(z^{-1})
\qquad (\epsilon_I \equiv 0).
\end{equation}
The situation changes drastically in an absorption band or in
the vicinity of it. From Eq.~(\ref{3.6}) it is
seen, that in an absorbing regime the
line width also becomes proportional to the inverse cube of the distance.
Moreover, the line shift can substantially increase. 

To demonstrate the possibility of improving the resolution 
of SNOM near a medium resonance,
we have calculated numerically the spontaneous decay rate as well as
the line shift using the single-resonance model permittivity
\begin{equation} \label{3.7}
\epsilon(\omega) =
1+\frac{(0.46\omega_T)^2}{\omega_T^2-\omega^2-i\gamma \omega} .
\end{equation}
\vspace*{-1ex}
\begin{figure}[h]
\begin{minipage}{0.5\textwidth}
\psfig{file=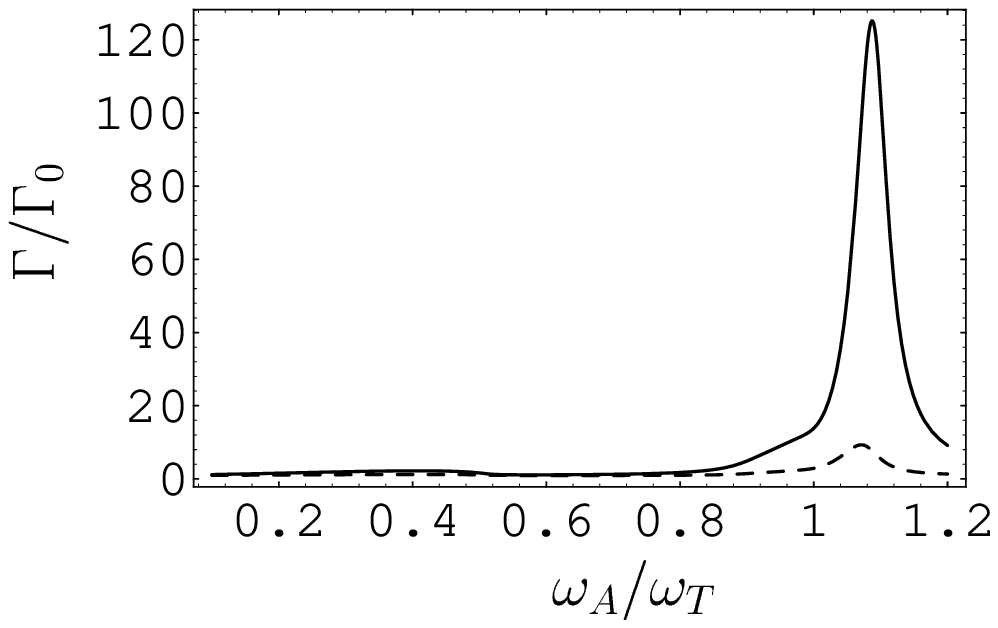,width=6.5cm}
\end{minipage}
\hfill
\begin{minipage}{0.5\textwidth}
\psfig{file=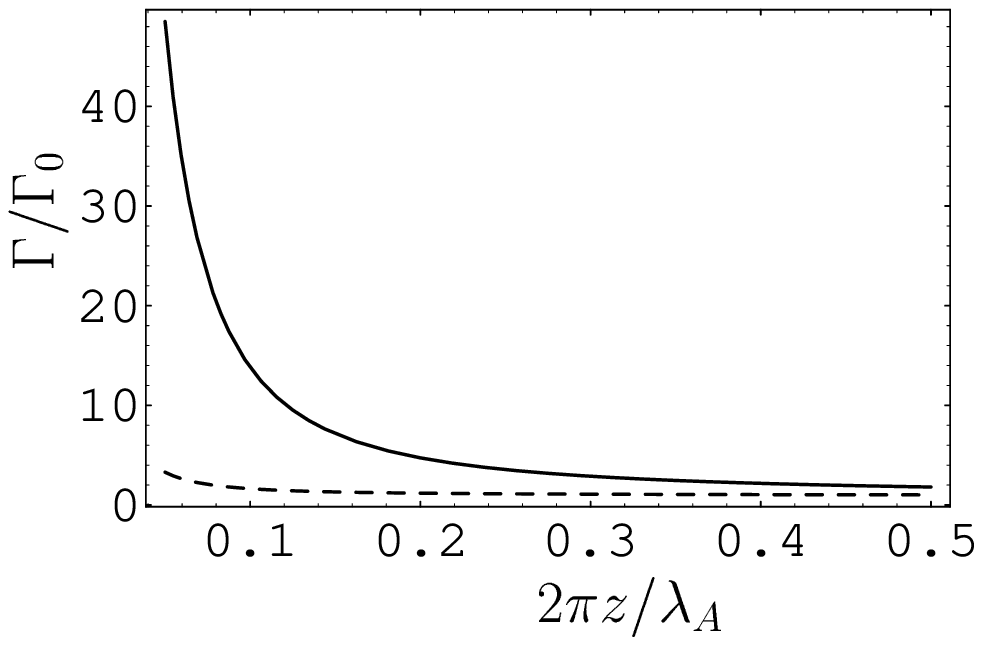,width=6.5cm}
\end{minipage}
\caption{\label{fig1} Fig.~1. Decay rate \protect$\Gamma/\Gamma_0$
(\protect$\mu$ $\!=$ $\mu_z$) for 
\protect$\gamma/\omega_A$ $\!=$ $\!0.05$  
[\protect$\kappa(\omega_A)$ $\!=$ $\!1.29$] as a function of frequency
(left figure) for \mbox{$2\pi z/\lambda_A$ $\!=$ $\!0.1$} (full line)
and $2\pi z/\lambda_A$ $\!=$ $\!0.3$ (dashed line), 
and as a function of distance (right figure) for
\protect$\omega_A/\omega_T$ $\!=$ $\!1$ (full line) and
\protect$\omega_A/\omega_T$ $\!=$ $\!0.5$ (dashed line).} 
\end{figure}
\vspace*{-4ex}
\begin{figure}[h]
\begin{minipage}{0.5\textwidth}
\psfig{file=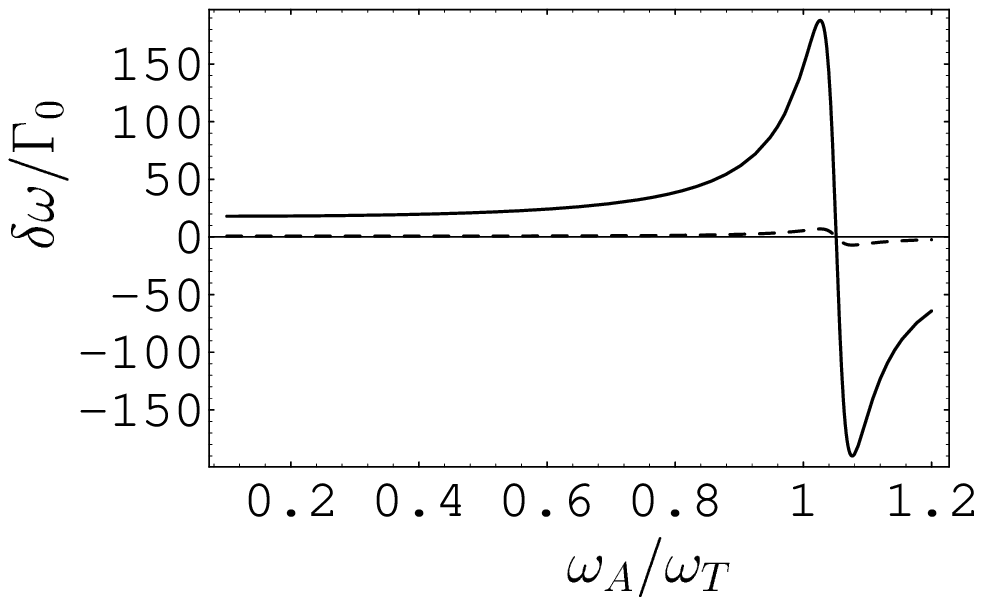,width=6.5cm}
\end{minipage}
\hfill
\begin{minipage}{0.5\textwidth}
\psfig{file=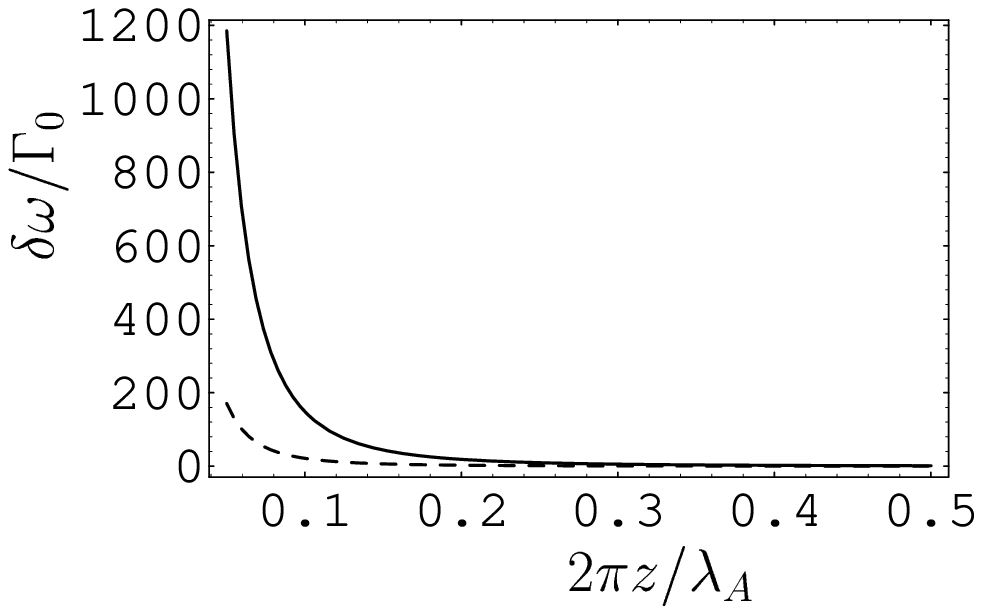,width=6.5cm}
\end{minipage}
\caption{\label{fig2} Fig.~2. Line shift
\protect$\delta\omega/\Gamma_0$ (\protect$\mu$ $\!=$ $\mu_z$) for
\protect$\gamma/\omega_A$ $\!=$ $\!0.05$ 
[\protect$\kappa(\omega_A)$ $\!=$ $\!1.29$] as a function of frequency
(left figure) for \mbox{$2\pi z/\lambda_A$ $\!=$ $\!0.1$} (full
line) and $2\pi z/\lambda_A$ $\!=$ $\!0.3$ (dashed line), and
as a function of distance (right figure) for
\protect$\omega_A/\omega_T$ $\!=$ $\!1$ (full line) and
\protect$\omega_A/\omega_T$ $\!=$ $\!0.5$ (dashed line).}
\end{figure}
   From Figures 1 and 2 one observes that  
near the medium resonance both the decay rate and the
line shift exceed substantially the values realized in standard 
SNOM. Moreover, the resolution of scanning
is improved, as can be seen from the
vertical line-width resolution
\vspace{-1ex}
\begin{equation}
\frac{\delta\Gamma}{\delta z} \propto \frac{1}{z^4}
\frac{\epsilon_I(\omega_A)}{|\epsilon(\omega_A)+1|^2} 
\end{equation}
(a similar formula holds for the vertical line-shift resolution).
Hence we are left with the proposal of
operating in an absorbing regime near a medium resonance in order to
improve the sensitivity of measurement.

%%%%%%%%%%%%%%%%%%%%%%%%%%%%%%%%%%%%%%%%%%%%%%%%%%%%%%%%%%%%%%%%%%%%%%

\section{An excited atom in a spherical cavity}
\label{real}

Another interesting geometry is the spherical one. In particular,
an atom in an empty sphere ${\cal V}_R$ of radius $R$ inside an otherwise
continuous dielectric medium can serve, in a sense, as a model for 
spontaneous decay in media (real-cavity model). 
For \mbox{${\bf r},{\bf r}'$ $\!\in$ $\!{\cal V}_R$} the Green function
can again be written in the form of Eq.~(\ref{3.1}), where the
reflection tensor reads in dyadic notation as [8]
\begin{eqnarray}
\label{4.0}
\lefteqn{
\mbox{\boldmath $R$}({\bf r},{\bf r'},\omega)  
=\frac{i\omega}{4\pi c} \sum\limits_{e,o}
\sum\limits_{n=1}^\infty \sum\limits_{m=0}^n \Bigg\{
\frac{2n+1}{n(n+1)} \frac{(n-m)!}{(n+m)!} (2-\delta_{0m}) 
}
\nonumber \\
&& \hspace*{-4ex} 
\times\!\left[ C_n^M(\omega)\mbox{\boldmath $M$}_{{e \atop o}nm}
\!\!\left( {\bf r},\frac{\omega}{c} \right) \!\mbox{\boldmath $M$}_{{e \atop o}nm}
\!\!\left( {\bf r'},\frac{\omega}{c} \right)\! +\!C_n^N(\omega)
\mbox{\boldmath $N$}_{{e \atop o}nm}\!\! \left( {\bf r},\frac{\omega}{c} \right)
\!\mbox{\boldmath $N$}_{{e \atop o}nm} \!\!\left( {\bf r'},\frac{\omega}{c} \right)
\right] \!\Bigg\} .
\qquad
\end{eqnarray}
Here, $\mbox{\boldmath $M$}_{{e \atop o}nm}({\bf r},k)$ 
and $\mbox{\boldmath $N$}_{{e \atop o}nm}({\bf r},k)$ are the 
($e$ven and $o$dd) vector Debye potentials, and the quantities 
$C_n^{M,N}(\omega)$ are the generalized reflection
coefficients. 
Combining Eqs.~(\ref{2.16a}) and (\ref{4.0}) (for ${\bf r}$
$\!=$ ${\bf r}'$ $\!=$ ${\bf r}_A$), after some algebra we 
derive 
\begin{equation} \label{4.1}
\Gamma = \left[ 1+ {\rm Re}\,C_1^N(\omega_A) \right] \Gamma_0,
\end{equation}
where
\begin{equation} \label{4.2}
C_1^N(\omega) = \frac{\left[i+z(n+1)-iz^2 n-z^3 n^2/(n+1)\right] 
e^{iz}}{\sin z-z(\cos z+in\sin z)+iz^2 n\cos z-z^3(\cos z-in\sin z)
n^2/(n^2-1)} 
\end{equation}
($z$ $\!=$ $\!R\omega/c$; for details, see [4]). Examples are
shown in Figs.~3 and 4. Note that the decay rate is solely related to
the transverse electromagnetic field, $\Gamma$ $\!=$ $\!\Gamma^\perp$. 

Restricting our attention to small cavities, 
$\omega_A R/c$ $\!\ll$ $\!1$, expansion yields
\begin{eqnarray}
\lefteqn{
\label{4.4}
\Gamma = \Bigg\{
\frac{9\epsilon_I(\omega_A)}{|2\epsilon(\omega_A)+1|^2} \left(
\frac{c}{\omega_A R} \right)^3 + \frac{9\epsilon_I(\omega_A) \left[
28|\epsilon(\omega_A)|^2 +12\epsilon_R(\omega_A)+1
\right]}{5|2\epsilon(\omega_A)+1|^4} \left( \frac{c}{\omega_A R}
\right) 
}
\nonumber \\ &&\hspace{2ex}
+\,\frac{9\eta(\omega_A)}{|2\epsilon(\omega_A)+1|^4} \bigg[
4|\epsilon(\omega_A)|^4 +4\epsilon_R(\omega_A) |\epsilon(\omega_A)|^2
+\epsilon_R^2(\omega_A) -\epsilon_I^2(\omega_A) \bigg] 
\hspace*{10ex}
\nonumber \\ &&\hspace{2ex} 
-\,\frac{9\kappa(\omega_A)\epsilon_I(\omega_A)}{|2\epsilon(\omega_A)+1|^4} 
\bigg[ 4|\epsilon(\omega_A)|^2 +2\epsilon_R(\omega_A) \bigg] \Bigg\} \Gamma_0
+{\cal O}(R)
\end{eqnarray}
($\epsilon_R$ $\!\equiv$ $\!{\rm Re}\,\epsilon$,
$\epsilon_I$ $\!\equiv$ $\!{\rm Im}\,\epsilon$).
When absorption is fully disregarded, i.e.,
${\rm Im}\,\epsilon$ $\!\equiv$ $\!0$, Eq.~(\ref{4.4})
reduces exactly to the well-known Glauber-Lewenstein formula [9]
\begin{equation} \label{4.3}
\Gamma =\left( \frac{3n^2}{2n^2+1} \right)^2 n \Gamma_0 \,.
\end{equation}
Needless to say that when setting $\epsilon$ $\!\equiv$ $\!1$, then the 
free-space spontaneous emission rate is recovered.
In absorbing media the decay rate sensitively depends on the 
cavity radius. Note that the terms $\sim R^{-3}$ and $\sim R^{-1}$  
result from absorption. In particular, the term
$\sim R^{-3}$ can be regarded as describing nonradiative energy 
transfer from the guest atom to the constituents of the dielectric 
via dipole-dipole interaction. It should be pointed out that
the above derived rate formula for absorbing media strongly 
contradicts the rate formula suggested in [10]. 

%%%%%%%%%%%%%%%%%%%%%%%%%%%%%%%%%%%%%%%%%%%%%%%%%%%%%%%%%%%%%%%%%%%%%%

\section{Virtual-cavity model of spontaneous decay}
\label{virtual}

In the (Clausius-Mosotti) virtual-cavity model it is assumed 
that the field outside the small sphere is not modified by the
sphere, and the local field inside is given by the
(smeared) macroscopic field corrected by adding a contribution
from the polarization of the sphere.
In quantum theory, one has to be careful with the definition of the 
local field, because of the nonvanishing noise polarization even
in the zero-temperature limit.  
In particular, simple multiplication of the macroscopic electric field 
with a correction factor as suggested in [10,11] leads to a wrong 
expression for the rate of spontaneous decay and 
misinterpretations concerning the action of transverse and longitudinal 
fields. Introducing the local field as 
$\underline{\hat{\bf E}}{}'({\bf r},\omega)$ $\!=$
$\underline{\hat{\bf E}}({\bf r},\omega)$ $\!+$
$\!\underline{\hat{\bf P}}({\bf r},\omega)/(3\epsilon_0)$,
from Eqs.~(\ref{2.2}) and (\ref{2.4}) together with Eqs.~(\ref{2.5}) 
-- (\ref{2.7}) it follows that [4,12]
\begin{equation}
\label{5.1}
\underline{\hat{\bf E}}{}'({\bf r},\omega) = \frac{1}{3} \big[
\epsilon({\bf r},\omega) +2 \big] \underline{\hat{\bf E}}({\bf
r},\omega) + \frac{1}{3\epsilon_0} \underline{\hat{\bf P}}{}^N({\bf
r},\omega), 
\end{equation}
with 
\begin{equation}
\underline{\hat{\bf P}}{}^N({\bf r},\omega) = -\frac{1}{i\omega}
\,\underline{\hat{\bf j}}({\bf r},\omega) =
i\sqrt{\frac{\hbar\epsilon_0}{\pi}\, \epsilon_I({\bf r},\omega)}
\,\hat{\bf f}({\bf r},\omega) .
\end{equation}

The spontaneous decay rate can now be calculated using
Eqs.~(\ref{2.15}) and (\ref{2.17}) together with the local field
defined in Eq.~(\ref{5.1}) and the Green function for bulk material,
$G^M_{kk'}({\bf r},{\bf r'},\omega_A)$. 
However, for ${\bf r}, {\bf r'}$ $\!\to$ $\!{\bf r}_A$ the 
imaginary part of $G^M_{kk'}({\bf r},{\bf r'},\omega_A)$ becomes singular,
which reflects the fact that the dielectric is thought of as
being smeared out even over the region where the atom is located. 
In order to overcome this difficulty, regularization is required. 
A small distance between the spatial arguments ${\bf r}$ and ${\bf r'}$ 
can be kept, and the decay rate can be obtained by averaging over an
appropriately chosen sphere of radius $R$ (virtual cavity), 
\begin{eqnarray} \label{5.2}
\Gamma &=& \frac{2\omega_A^2 \mu_k \mu_{k'}}{\hbar \epsilon_0 c^2} \left|
\frac{\epsilon(\omega_A)+2}{3} \right|^2 {\rm Im} \,
\overline{G^M_{kk'}({\bf r},{\bf r'},\omega_A)}
+\frac{2\mu_k \mu_{k'}}{9\hbar \epsilon_0} \epsilon_I(\omega_A) 
\,\overline{\delta_{kk'} \delta({\bf r}-{\bf r'})} \nonumber \\ &&
+\frac{4\omega_A^2\mu_k \mu_{k'}}{3\hbar \epsilon_0 c^2} \,\epsilon_I(\omega_A) 
\,{\rm Re} \left[ \frac{\epsilon(\omega_A)+2}{3}
\,\overline{G^M_{kk'}({\bf r},{\bf r'},\omega_A)} \right] 
\end{eqnarray}
(for details, see [4]). Obviously, different averaging procedures  
leads to (slightly) different numerical prefactors in 
the $R$-dependent terms [4,12]. 

Let us consider a small cavity ($|R\omega_A n/c|$ $\!\ll$ $\!1$). 
Averaging with respect to both spatial arguments separately over the
sphere with equal weight, we obtain 
[12]
\begin{equation}
\label{5.2a}
\Gamma = \Gamma^\perp + \Gamma^\|,
\end{equation}
\begin{eqnarray}
\label{5.3}
\lefteqn{
\hspace{-6ex} 
\Gamma^\perp = \Bigg\{ \frac{25\epsilon_I(\omega_A)}{54} \left(
\frac{c}{\omega_A R} \right)^3 +\epsilon_I(\omega_A) \left[
\epsilon_R(\omega_A) +2 \right] \left[ \frac{8}{15} \left(
\frac{c}{\omega_A R} \right) -\frac{2\kappa(\omega_A)}{9} \right]
}
\nonumber \\ && \hspace{10ex} 
+\,\eta(\omega_A) \left[ \left|
\frac{\epsilon(\omega_A)+2}{3} \right|^2
-\frac{2\epsilon_I^2(\omega_A)}{9} \right]\Bigg\}\Gamma_0 
+{\cal O}(R) ,
\end{eqnarray}
\begin{equation}
\label{5.4}
\Gamma^\| = \Gamma_0
\frac{4\epsilon_I(\omega_A)}{27|\epsilon(\omega_A)|^2}
\left( \frac{c}{\omega_A R} \right)^3 .
\end{equation}
Here, $\Gamma^\perp$ and $\Gamma^\|$, respectively, are related to the 
transverse and longitudinal parts of the electromagnetic fields. 
It should be pointed out that the appearance of $\Gamma^\|$
is due to the fact that (in contrast of the real-cavity model) the 
atom is not strictly located in vacuum and
the imaginary part of the longitudinal Green function
(at equal space points) does not vanish for absorbing media. 
\begin{figure}[h]
\begin{minipage}{0.5\textwidth}
\psfig{file=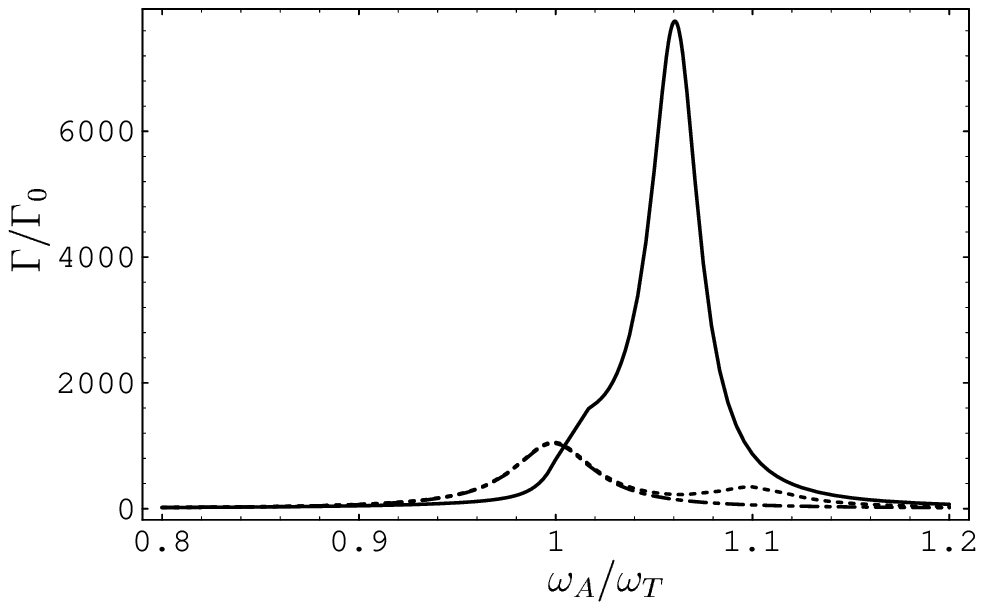,width=6.5cm}
\end{minipage}
\hfill
\begin{minipage}{0.5\textwidth}
\psfig{file=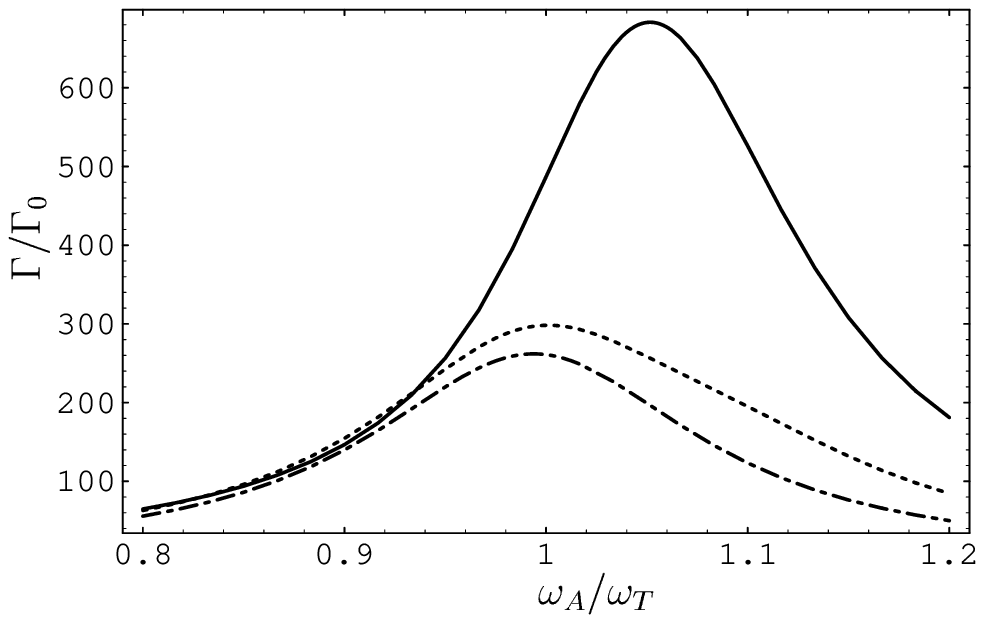,width=6.5cm}
\end{minipage}
\caption{Fig.~3. Spontaneous decay rate \protect$\Gamma/\Gamma_0$ for $R$ $\!=$
$\!0.02\lambda_A$ and $\gamma$ $\!=$ $\!0.05\omega_T$ (left figure)
and $\gamma$ $\!=$ $\!0.2\omega_T$ (right figure). 
The solid lines correspond to the real-cavity model, Eq.~(\protect\ref{4.4}),
and the dotted lines correspond to the virtual-cavity model,
Eqs.~(\protect\ref{5.3}) and (\protect\ref{5.4}), with the broken line 
indicating the transverse part (\protect\ref{5.3}).}
\end{figure}
\vspace*{-2ex}
\begin{figure}[h]
\begin{minipage}{0.5\textwidth}
\psfig{file=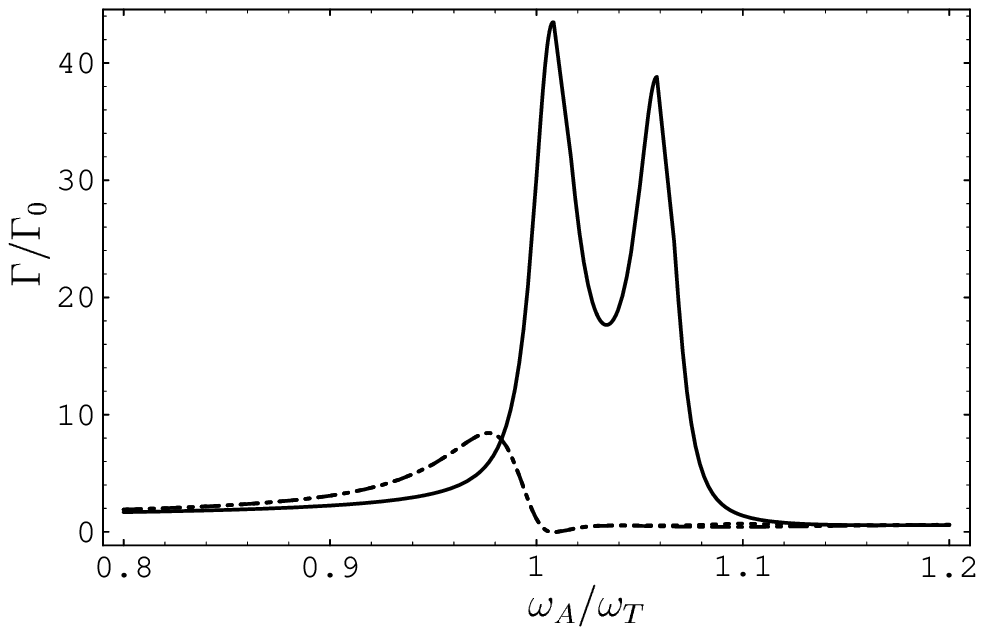,width=6.5cm}
\end{minipage}
\hfill
\begin{minipage}{0.5\textwidth}
\psfig{file=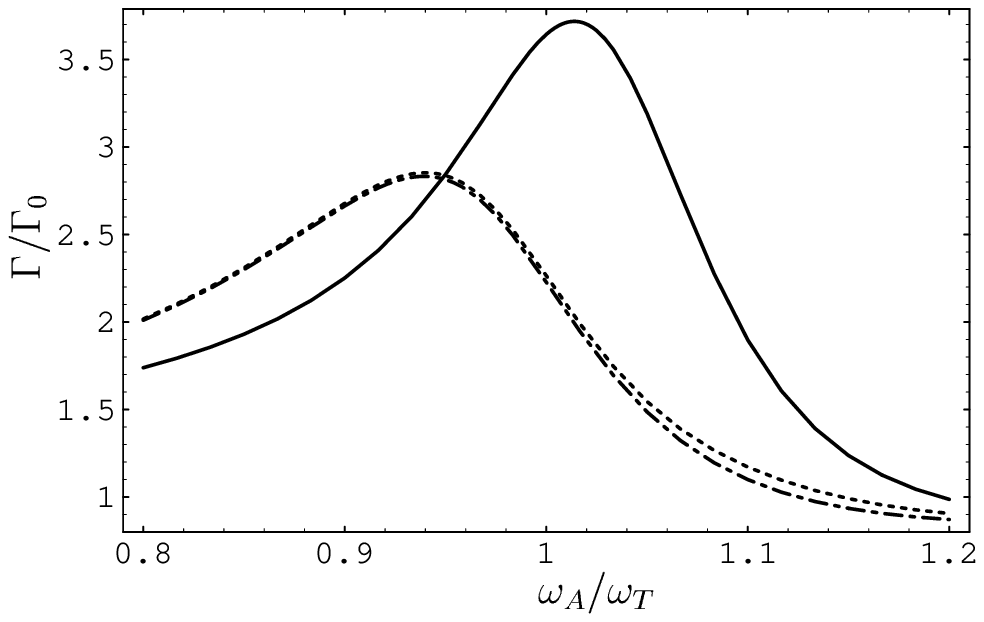,width=6.5cm}
\end{minipage}
\caption{Fig.~4. Spontaneous decay rate \protect$\Gamma/\Gamma_0$ for $R$ $\!=$
$\!0.2\lambda_A$ and $\gamma$ $\!=$ $\!0.05\omega_T$ (left figure)
and $\gamma$ $\!=$ $\!0.2\omega_T$ (right figure). 
The solid lines correspond to the real-cavity model, Eq.~(\protect\ref{4.4}),
and the dotted lines correspond to the virtual-cavity model,
Eqs.~(\protect\ref{5.3}) and (\protect\ref{5.4}), with the broken line 
indicating the transverse part (\protect\ref{5.3}).}
\end{figure}

Examples of $\Gamma$ and $\Gamma^\perp$ are
shown in Figs.~3 and 4. The results differ from those in [10,11] in 
the contribution of the noise polarization $\hat{\bf P}^N$, which 
is disregarded in [10,11]. It is worth noting that the effect
of the noise polarization has recently been fully confirmed
on the basis of a microscopic approach [13]. 
The effect vanishes for nonabsorbing media ($\epsilon_I$ $\!\equiv$ $\!0$), 
and Eq.~(\ref{5.3}) yields the well-known rate formula
\begin{equation} \label{5.5}
\Gamma = \Gamma^\perp = \left( \frac{n^2+2}{3}
\right)^2  n \Gamma_0. 
\end{equation}

%%%%%%%%%%%%%%%%%%%%%%%%%%%%%%%%%%%%%%%%%%%%%%%%%%%%%%%%%%%%%%%%%%%%%%

\section{Summary}
\label{summary}

We have shown how quantization of the phenomenological electrodynamics
in linear media can be most easily employed to study radiating
atoms in the presence of dispersive and absorbing dielectric bodies
and to calculate the line widths and line shifts of the atomic 
transitions. It is worth noting that the approach reproduces typical 
results that have been derived using quite different concepts
including involved microscopic ones. The approach also shows that
the rate formulas given in [10,11] for the spontaneous decay 
of excited atoms in absorbing media are not correct, because they 
are based on wrong respective incomplete medium-assisted vacuum noise.
In particular, we have considered a planar interface and a
spherical cavity, which may serve as a model of spontaneous 
decay in media, and we have compared it with the corresponding 
(Clausius-Mosotti) virtual-cavity model. 
The results reveal that when an excited atom approaches
an absorbing dielectric body, then the rate of spontaneous decay 
becomes proportional to the inverse cubic distance of the atom from  
dielectric body [see, e.g., Eqs.~(\ref{3.6}) and (\ref{4.4})].
In microscopic models, terms of that type have been shown to correspond 
to nonradiative energy transfer from the atom to the medium via dipole-dipole
interaction, i.e., a second-order process with no photon in the final state. 
With regard to applications, the results show that absorption 
offers the possibility of improving the resolution of SNOM.

%%%%%%%%%%%%%%%%%%%%%%%%%%%%%%%%%%%%%%%%%%%%%%%%%%%%%%%%%%%%%%%%%%%%%%

\medskip

\noindent {\bf Acknowledgements}
This work was supported by the Deutsche Forschungsgemeinschaft.
We thank M. Fleischhauer, 
C. Henkel, and D. Meschede for
helpful and stimulating discussions.

\small
\kapitola{References}
\begin{description}
\itemsep0pt
\item{\ [1]} \refer{T. Gruner, D.-G. Welsch}{Phys. Rev. A}{53}{1996}{1818}
\item{\ [2]} \refer{Ho Trung Dung, L. Kn\"oll,
D.-G. Welsch}{Phys. Rev. A}{57}{1998}{3931} 
\item{\ [3]} \refer{S. Scheel, L. Kn\"oll,
D.-G. Welsch}{Phys. Rev. A}{58}{1998}{700}
\item{\ [4]} S. Scheel, L. Kn\"oll, D.-G. Welsch:
``{\sl Spontaneous decay of an excited atom in an absorbing
dielectric}'',
submitted to {\sl Phys. Rev. A} [quant-ph/9904015 (1999)]; 
\item{\ [5]} \refer{M. Fichet, F. Schuller, D. Bloch,
M. Ducloy}{Phys. Rev. A}{51}{1995}{1553}
\item{\ [6]} \refer{C. Henkel, V. Sandoghdar}{Opt. Comm.}{158}{1998}{250}
\item{\ [7]} \refer{M.S. Yeung, T.K. Gustafson}{Phys. Rev. A}{54}{1996}{5227}
\item{\ [8]} \refer{L.W. Li, P.S. Kooi, M.S. Leong, 
T.S. Yeo}{IEEE Transactions on Microwave Theory and
Techniques}{42}{1994}{2302}
\item{\ [9]} \refer{R.J. Glauber,
M. Lewenstein}{Phys. Rev. A}{43}{1991}{467} 
\item{[10]} \refer{S.M. Barnett, B. Huttner, R. Loudon,
R. Matloob}{J. Phys. B: At. Mol. Opt. Phys.}{29}{1996}{3763}
\item{[11]}  \refer{S.M. Barnett, B. Huttner,
R. Loudon}{Phys. Rev. Lett.}{68}{1992}{3698} 
\item{[12]}  S. Scheel, L. Kn\"oll, D.-G. Welsch, S.M. Barnett:
``{\sl Quantum local-field correction and spontaneous decay}'',
submitted to {\sl Phys. Rev. A} [quant-ph/9811067 (1998)];
\item{[13]}  \refer{M. Fleischhauer,
S.F. Yelin}{Phys. Rev. A}{59}{1999}{2427} 
M. Fleischhauer: ``{\sl Spontaneous emission and level shifts in 
absorbing disordered dielectrics and dense atomic gases: 
A Green's function approach}''
[quant-ph/9902076 (1999)]. 
\end{description}

\end{document}